\documentclass[prl,twocolumn,superscriptaddress,preprintnumbers,amsmath,amssymb]{revtex4}
\usepackage{graphicx}
\usepackage{amsmath}
\usepackage{latexsym}
\usepackage{dcolumn}
\usepackage{bm}
\usepackage{float}
\usepackage{graphicx,here}
\usepackage{color}

\emergencystretch=\maxdimen
\begin{document}

\title{\bf Observation of quantum spin Hall states in Ta$_2$Pd$_3$Te$_5$}

\affiliation{Institute of Physics, Chinese Academy of Sciences, Beijing 100190, China}
\affiliation{Hiroshima Synchrotron Radiation Center, Hiroshima University, 2-313 Kagamiyama, Higashi-Hiroshima 739-0046, Japan}
\affiliation{Department of Materials Science and Engineering, Stanford University, Stanford, California 94305, USA}
\affiliation{School of Physical Sciences, University of Chinese Academy of Sciences, Beijing 100049, China}
\affiliation{Center of Materials Science and Optoelectronics Engineering, University of Chinese Academy of Sciences, Beijing 100049, China}
\affiliation{Songshan Lake Materials Laboratory, Dongguan, Guangdong 523808, China}
\affiliation{These authors contributed equally to this work.}

\author{Xuguang Wang}
\affiliation{Institute of Physics, Chinese Academy of Sciences, Beijing 100190, China}
\affiliation{School of Physical Sciences, University of Chinese Academy of Sciences, Beijing 100049, China}
\affiliation{These authors contributed equally to this work.}
\author{Daiyu Geng}
\affiliation{Institute of Physics, Chinese Academy of Sciences, Beijing 100190, China}
\affiliation{School of Physical Sciences, University of Chinese Academy of Sciences, Beijing 100049, China}
\affiliation{These authors contributed equally to this work.}
\author{Dayu Yan}
\affiliation{Institute of Physics, Chinese Academy of Sciences, Beijing 100190, China}
\affiliation{School of Physical Sciences, University of Chinese Academy of Sciences, Beijing 100049, China}
\affiliation{These authors contributed equally to this work.}
\author{Wenqi Hu}
\affiliation{Institute of Physics, Chinese Academy of Sciences, Beijing 100190, China}
\affiliation{School of Physical Sciences, University of Chinese Academy of Sciences, Beijing 100049, China}
\affiliation{These authors contributed equally to this work.}
\author{Hexu Zhang}
\affiliation{Institute of Physics, Chinese Academy of Sciences, Beijing 100190, China}
\affiliation{School of Physical Sciences, University of Chinese Academy of Sciences, Beijing 100049, China}
\author{Shaosheng Yue}
\affiliation{Institute of Physics, Chinese Academy of Sciences, Beijing 100190, China}
\affiliation{School of Physical Sciences, University of Chinese Academy of Sciences, Beijing 100049, China}
\author{Zhenyu Sun}
\affiliation{Institute of Physics, Chinese Academy of Sciences, Beijing 100190, China}
\affiliation{School of Physical Sciences, University of Chinese Academy of Sciences, Beijing 100049, China}
\author{Shiv Kumar}
\affiliation{Hiroshima Synchrotron Radiation Center, Hiroshima University, 2-313 Kagamiyama, Higashi-Hiroshima 739-0046, Japan}
\author{Kenya Shimada}
\affiliation{Hiroshima Synchrotron Radiation Center, Hiroshima University, 2-313 Kagamiyama, Higashi-Hiroshima 739-0046, Japan}
\author{Peng Cheng}
\affiliation{Institute of Physics, Chinese Academy of Sciences, Beijing 100190, China}
\affiliation{School of Physical Sciences, University of Chinese Academy of Sciences, Beijing 100049, China}
\author{Lan Chen}
\affiliation{Institute of Physics, Chinese Academy of Sciences, Beijing 100190, China}
\affiliation{School of Physical Sciences, University of Chinese Academy of Sciences, Beijing 100049, China}
\affiliation{Songshan Lake Materials Laboratory, Dongguan, Guangdong 523808, China}
\author{Simin Nie}
\affiliation{Department of Materials Science and Engineering, Stanford University, Stanford, California 94305, USA}
\author{Zhijun Wang}
\affiliation{Institute of Physics, Chinese Academy of Sciences, Beijing 100190, China}
\affiliation{School of Physical Sciences, University of Chinese Academy of Sciences, Beijing 100049, China}
\author{Youguo Shi}\thanks{ygshi@iphy.ac.cn}
\affiliation{Institute of Physics, Chinese Academy of Sciences, Beijing 100190, China}
\affiliation{School of Physical Sciences, University of Chinese Academy of Sciences, Beijing 100049, China}
\affiliation{Center of Materials Science and Optoelectronics Engineering, University of Chinese Academy of Sciences, Beijing 100049, China}
\affiliation{Songshan Lake Materials Laboratory, Dongguan, Guangdong 523808, China}
\author{Yi-Qi Zhang}
\affiliation{Institute of Physics, Chinese Academy of Sciences, Beijing 100190, China}
\affiliation{School of Physical Sciences, University of Chinese Academy of Sciences, Beijing 100049, China}
\author{Kehui Wu}\thanks{khwu@iphy.ac.cn}
\affiliation{Institute of Physics, Chinese Academy of Sciences, Beijing 100190, China}
\affiliation{School of Physical Sciences, University of Chinese Academy of Sciences, Beijing 100049, China}
\affiliation{Songshan Lake Materials Laboratory, Dongguan, Guangdong 523808, China}
\author{Baojie Feng}\thanks{bjfeng@iphy.ac.cn}
\affiliation{Institute of Physics, Chinese Academy of Sciences, Beijing 100190, China}
\affiliation{School of Physical Sciences, University of Chinese Academy of Sciences, Beijing 100049, China}

\date{\today}

\clearpage

\begin{abstract}
Two-dimensional topological insulators (2DTIs), which host the quantum spin Hall (QSH) effect, are one of the key materials in next-generation spintronic devices. To date, experimental evidence of the QSH effect has only been observed in a few materials, and thus, the search for new 2DTIs is at the forefront of physical and materials science. Here, we report experimental evidence of a 2DTI in the van der Waals material Ta$_2$Pd$_3$Te$_5$. First-principles calculations show that each monolayer of Ta$_2$Pd$_3$Te$_5$ is a 2DTI with weak interlayer interactions. Combined transport, angle-resolved photoemission spectroscopy, and scanning tunneling microscopy measurements confirm the existence of a band gap at the Fermi level and topological edge states inside the gap. These results demonstrate that Ta$_2$Pd$_3$Te$_5$ is a promising material for fabricating spintronic devices based on the QSH effect.
\end{abstract}

\maketitle

Two-dimensional topological insulators (2DTIs), also known as quantum spin Hall (QSH) insulators, feature a bulk band gap and helical in-gap states at the material boundaries \cite{HasanMZ2010,QiXL2011,AndoY2013}. The edge states of a 2DTI can serve as one-dimensional conducting channels in which backscattering is forbidden by time-reversal symmetry. Therefore, 2DTIs provide an ideal platform to fabricate low-dissipation spintronic devices. To date, 2DTIs have only been realized in two types of materials. The first type is quantum well systems, including HgTe/CdHgTe \cite{KonigM2007} and InAs/GaSb \cite{LiuC2008,KnezI2011,DuL2015}. However, the material synthesis of these quantum wells is extremely challenging, and experimental signatures of the QSH effect have only been observed by a few research groups. The second type of 2DTI has been reported in several two-dimensional (2D) materials, such as bilayer Bi \cite{YangF2012,DrozdovIK2014}, monolayer 1T$^\prime$ WTe$_2$ \cite{TangS2017,JiaZ2017,PengL2017}, and graphene-like honeycomb lattices \cite{ZhuF2015,ReisF2017,DengJ2018,XuC2018,ZhuSY2019}. However, despite the large number of 2D materials as candidate 2DTIs, transport evidence of the QSH effect has only been reported in monolayer 1T$^\prime$ WTe$_2$ \cite{FeiZ2017,WuS2018,ShiY2019}.

Realizing QSH states in van der Waals materials, such as WTe$_2$, offers great opportunities to fabricate quantum transport devices, as monolayer or few-layer materials for realizing the QSH effect are very easy to obtain. However, as a prototypical van der Waals material, 1T$^\prime$ WTe$_2$ was confirmed to be a 2DTI only in the monolayer limit \cite{WuS2018}. In bulk form, WTe$_2$ becomes a metal with zero energy gap \cite{AliMN2014,SoluyanovAA2015}. As a result, QSH states disappear in multilayer WTe$_2$ because of the hybridization of the bulk and edge states. Therefore, QSH devices based on WTe$_2$ suffer from easy degradation of monolayer samples under ambient conditions. Since multilayer materials are typically more inert and have higher tunability via the thickness or twist angles, realizing QSH states in multilayer materials is highly desirable. This requires a semiconducting van der Waals material that hosts a similar inverted gap as the monolayer.

\begin{figure*}[htb]
\centering
\includegraphics[width=17 cm]{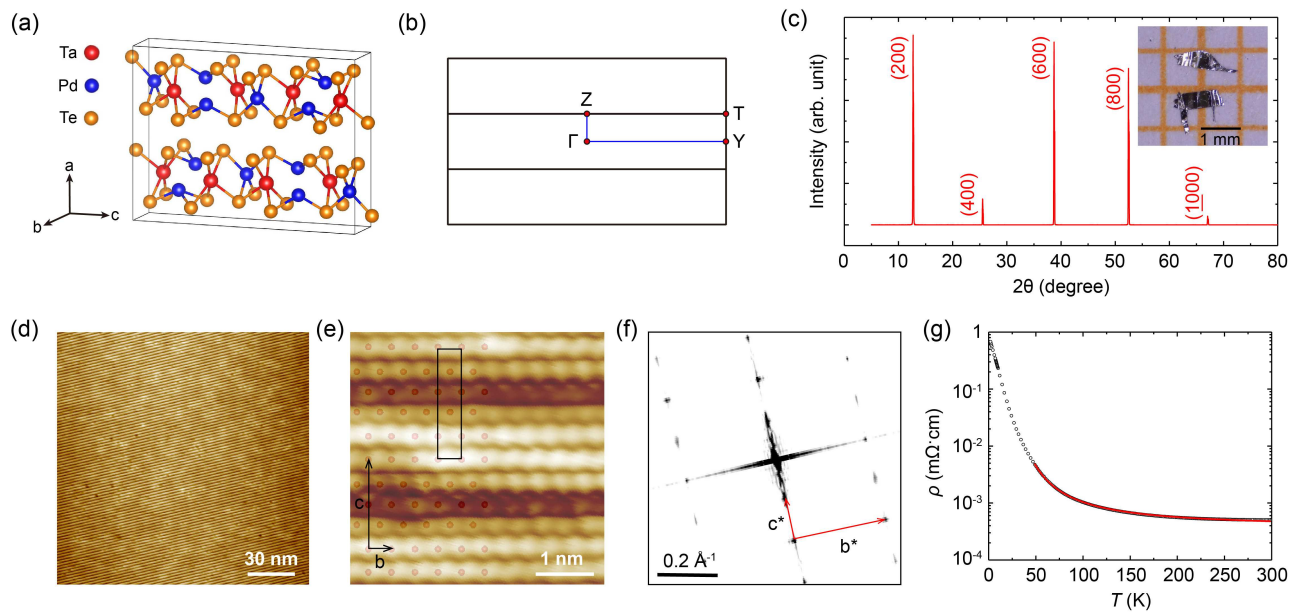}
\caption{{\bf Structure and characterization of Ta$_2$Pd$_3$Te$_5$ single crystals.} (a) Schematic drawing of the atomic structure of Ta$_2$Pd$_3$Te$_5$. (b) Schematic drawing of the Brillouin zones and high-symmetry points of monolayer Ta$_2$Pd$_3$Te$_5$. (c) X-ray diffraction spectrum on a flat surface of Ta$_2$Pd$_3$Te$_5$. The inset shows a photograph of typical Ta$_2$Pd$_3$Te$_5$ single crystals. (d) Large-scale STM topographic image of the Ta$_2$Pd$_3$Te$_5$ surface ($V\rm_B$=1 V; $I$=0.05 nA). (e) Zoomed-in STM image showing atomic resolution ($V\rm_B$=50 mV; $I$=0.05 nA). The atomic structure of the surface Te atoms (red balls) is superimposed on the left part of the image. (f) Fast Fourier transformed STM image, which shows a rectangular reciprocal lattice. Red arrows indicate the reciprocal lattice vectors. (g) In-plane resistivity of Ta$_2$Pd$_3$Te$_5$ single crystals as a function of temperature. The red line is the fitting result obtained using the Arrhenius formula.}
\end{figure*}

In this work, we report the observation of QSH states in the van der Waals material Ta$_2$Pd$_3$Te$_5$ \cite{NieS2020}, which hosts a band gap in both the bulk and monolayer forms. We synthesize Ta$_2$Pd$_3$Te$_5$ single crystals and investigate their electronic structures by combined first-principles calculations, transport, angle-resolved photoemission spectroscopy (ARPES), and scanning tunneling microscopy/spectroscopy (STM/STS) measurements. We prove that Ta$_2$Pd$_3$Te$_5$ hosts a band gap at the Fermi level. Because of the weak interlayer coupling, the topmost layer of Ta$_2$Pd$_3$Te$_5$ can be viewed as a monolayer material placed on a single-crystal substrate. As expected, we directly observe topological edge states using STS. These results provide strong evidence for QSH states in Ta$_2$Pd$_3$Te$_5$. The discovery of QSH states in van der Waals materials with a significant band gap could pave the way to realizing practical QSH devices.

Ta$_2$Pd$_3$Te$_5$ single crystals were synthesized by the self-flux method. The starting materials of Ta (99.999\%), Pd (99.9999\%), and Te (99.9999\%) were mixed in an Ar-filled glove box at a molar ratio of Ta:Pd:Te=2:4.5:7.5. The mixture was placed in an alumina crucible and sealed in an evacuated quartz tube. The tube was heated to 950 $^{\circ}$C over 10 h and maintained at this temperature for 2 days. Then, the tube was slowly cooled to 800 $^{\circ}$C at a rate of 0.5 $^{\circ}$C/h. Finally, the extra flux was removed by centrifugation at 800 $^{\circ}$C. After centrifugation, single crystals of Ta$_2$Pd$_3$Te$_5$ could be selected from the remnants in the crucible. To investigate the crystalline structure, single-crystal X-ray diffraction (XRD) was carried out at 273 K using Mo $\emph{K}\alpha$ radiation ($\lambda$ = 0.71073 {\AA}). The crystalline structure was refined by the full-matrix least-squares method on \emph{F$^2$} by using the SHELXL-2018/3 program. Electrical resistivity (\emph{$\rho$}) measurements were carried out on a physical property measurement system (PPMS, Quantum Design Inc.) using a standard dc four-probe technique.

ARPES experiments were performed at beamline BL-1 of the Hiroshima synchrotron radiation center \cite{IwasawaH2017}. The clean surfaces required for the ARPES measurements were obtained by cleaving the samples in an ultrahigh vacuum chamber with a base pressure of 1.0$\times$10$^{-9}$ Pa. Both the cleavage process and ARPES measurements were performed at 30 K. The energy resolution of the ARPES measurements was approximately 15 meV. STM/STS experiments were carried out in a home-built low-temperature ($\sim$5 K) STM system with a base pressure of 2$\times$10$^{-8}$ Pa. The clean surfaces for STM/STS measurements were also obtained by cleaving the samples {\it in situ} at low temperature.

First-principles calculations were performed within the framework of the projector augmented wave (PAW) method \cite{Blochl1994,Kresse1999} implemented in the Vienna {\it ab initio} simulation (VASP) package \cite{KresseG1996,KresseG1996prb}. The Perdew-Burke-Ernzerhof (PBE) generalized gradient approximation (GGA) exchange-correlation functional \cite{PerdewJP1996} was implemented in the calculations. The cutoff energy for plane wave expansion was 500 eV. Spin-orbit coupling was self-consistently taken into account within the second variational method. A 4-unit-cell slab structure (with 20 \AA{} vacuum) was built to simulate the surface spectrum.

Ta$_2$Pd$_3$Te$_5$ crystallizes in an orthorhombic structure with the space group \emph{Pnma} (No. 62). Schematic drawings of the atomic structure and Brillouin zones (BZs) of Ta$_2$Pd$_3$Te$_5$ are shown in Figs. 1(a) and 1(b), respectively. Each unit cell contains two Ta$_2$Pd$_3$Te$_5$ monolayers, which are stacked along the $a$ direction via weak van der Waals interactions. Each monolayer contains a Ta-Pd mixed layer sandwiched between two Te layers. Figure 1(c) shows the XRD spectrum on a flat surface of Ta$_2$Pd$_3$Te$_5$, whereby only $(h00)$ peaks are observed. A photograph of a typical Ta$_2$Pd$_3$Te$_5$ crystal is displayed in the inset of Fig. 1(c). The picture shows that the crystal is as large as 1 mm and has shiny surfaces, indicating the high crystallinity of our samples. The lattice parameters of Ta$_2$Pd$_3$Te$_5$ determined from the XRD data are \emph{a} = 13.9531(6) {\AA}, \emph{b} = 3.7038(2) {\AA}, and \emph{c} = 18.5991(8) {\AA}. Figure 1(d) shows a large-scale STM image of Ta$_2$Pd$_3$Te$_5$ (the \emph{bc} plane). The surface is slightly corrugated, forming periodic stripes along the \emph{b} direction. A zoomed-in STM image with atomic resolution is displayed in Fig. 1(e). Each bright protrusion corresponds to a Te atom, which well matches the structure model of Ta$_2$Pd$_3$Te$_5$. From the STM image, the rectangular structure of the \emph{bc} plane of Ta$_2$Pd$_3$Te$_5$ can also be identified in the fast Fourier transformed image in Fig. 1(e). The lattice constants determined from our STM results are $\sim$3.6 \AA{} and $\sim$18.7 \AA, respectively, which agree well with the lattice constants along the $b$ and $c$ directions.

The temperature dependence of the electrical resistivity of Ta$_2$Pd$_3$Te$_5$ is displayed in Fig. 1(g). When the temperature is decreased from 300 K to 2 K, the resistivity increases monotonically, indicating semiconductor behavior. The temperature-dependent resistivity can be fitted with the Arrhenius model $\rho$ $\sim$ exp($\epsilon_{act}$/$k_B$$T$), where $k_B$ and $\epsilon_{act}$ are the Boltzmann constant and thermal activation energy, respectively. The fitting results are shown by the red line in Fig. 1(g). The fitted $\epsilon_{act}$ is approximately 14 meV. Therefore, bulk Ta$_2$Pd$_3$Te$_5$ is a narrow-gap semiconductor with a global band gap of $\sim$14 meV.

\begin{figure}[htb]
\centering
\includegraphics[width=8.5 cm]{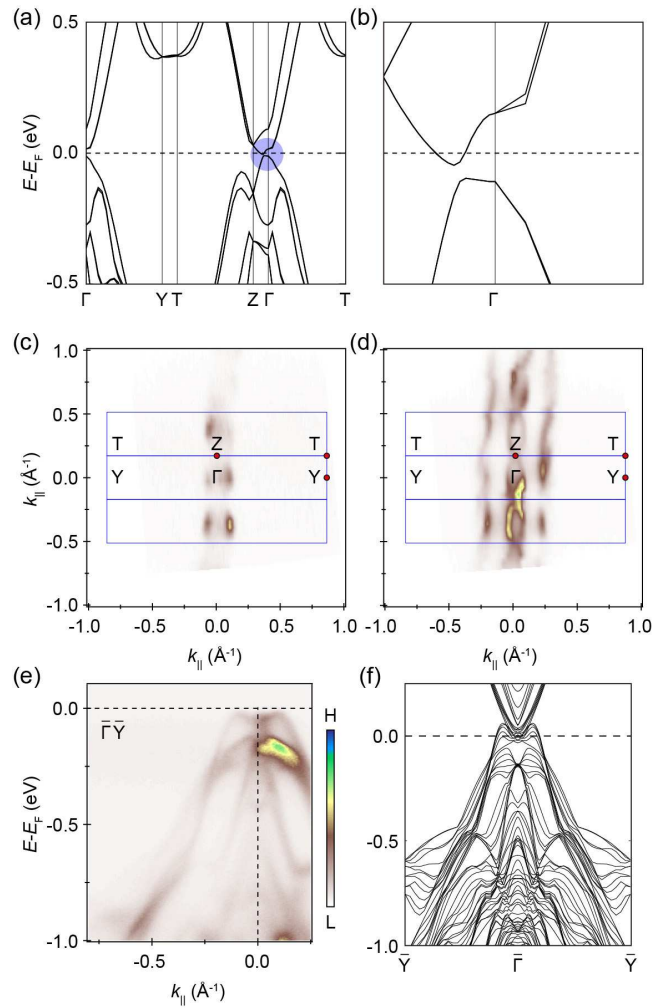}
\caption{{\bf Topological band structure of Ta$_2$Pd$_3$Te$_5$.} (a) Calculated band structure of monolayer Ta$_2$Pd$_3$Te$_5$. (b) Magnified view of the blue-shaded area in (a), showing the existence of a band gap. The calculated gap is approximately 5 meV. (c) and (d) ARPES intensity plots at the Fermi level and E$_B$=0.35 eV, respectively. The blue lines indicate the surface BZs of Ta$_2$Pd$_3$Te$_5$. (e) ARPES intensity plot of the band structure along the $\bar{\Gamma}$--$\rm\bar{Y}$ direction. (f) Slab calculation results of the band structure along the $\bar{\Gamma}$--$\rm\bar{Y}$ direction.}
\end{figure}

Before showing further experimental results of Ta$_2$Pd$_3$Te$_5$, we briefly discuss the topological properties based on our first-principles calculation results. For bulk Ta$_2$Pd$_3$Te$_5$, the symmetry indicators ($\mathbb Z_2\times \mathbb Z_2\times \mathbb Z_2\times \mathbb Z_4$) are 0. However, it has a nontrivial mirror Chern number in the $k_y=0$ plane due to the band inversion at the $\Gamma$ point \cite{NieS2020}, which indicates the topological nature of bulk Ta$_2$Pd$_3$Te$_5$. In the monolayer limit, Ta$_2$Pd$_3$Te$_5$ becomes a 2DTI \cite{NieS2020} with a similar band inversion. Its nontrivial topology has also been confirmed by the one-dimensional Wilson loop method. Figure 2(a) shows the band structure of monolayer Ta$_2$Pd$_3$Te$_5$, where an inverted band gap near the Fermi level can be identified. The calculated gap along the $\Gamma$--Z direction is approximately 5 meV. Notably, the gap fitted based on our transport measurements ($\sim$14 meV) on bulk samples is larger than the calculated value. We will later show that our STS measurements also indicate a significantly larger gap compared to the calculation results. This inconsistency probably originates from the fact that density functional theory (DFT) calculations may underestimate the band gap of materials.

To study the electronic structure of Ta$_2$Pd$_3$Te$_5$, we performed ARPES measurements on a freshly cleaved surface. An ARPES intensity map of the Fermi surface is displayed in Fig. 2(c), which shows a weak spectral weight along the $\bar{\Gamma}$--$\rm\bar{Y}$ direction. Because of the large lattice constant along the \emph{c} direction, we observed four BZs along $\bar{\Gamma}$--$\rm\bar{Z}$. With increasing binding energy, an oval-like pocket appears at the $\bar{\Gamma}$ point, as shown in Fig. 2(d). The band structure along the $\bar{\Gamma}$--$\rm\bar{Y}$ direction is shown in Fig. 2(e), which agrees well with our slab calculation results (see Fig. 2(f)). These ARPES results, combined with the DFT calculations, provide strong evidence of the topological band structure of Ta$_2$Pd$_3$Te$_5$.

\begin{figure*}[htb]
\centering
\includegraphics[width=15 cm]{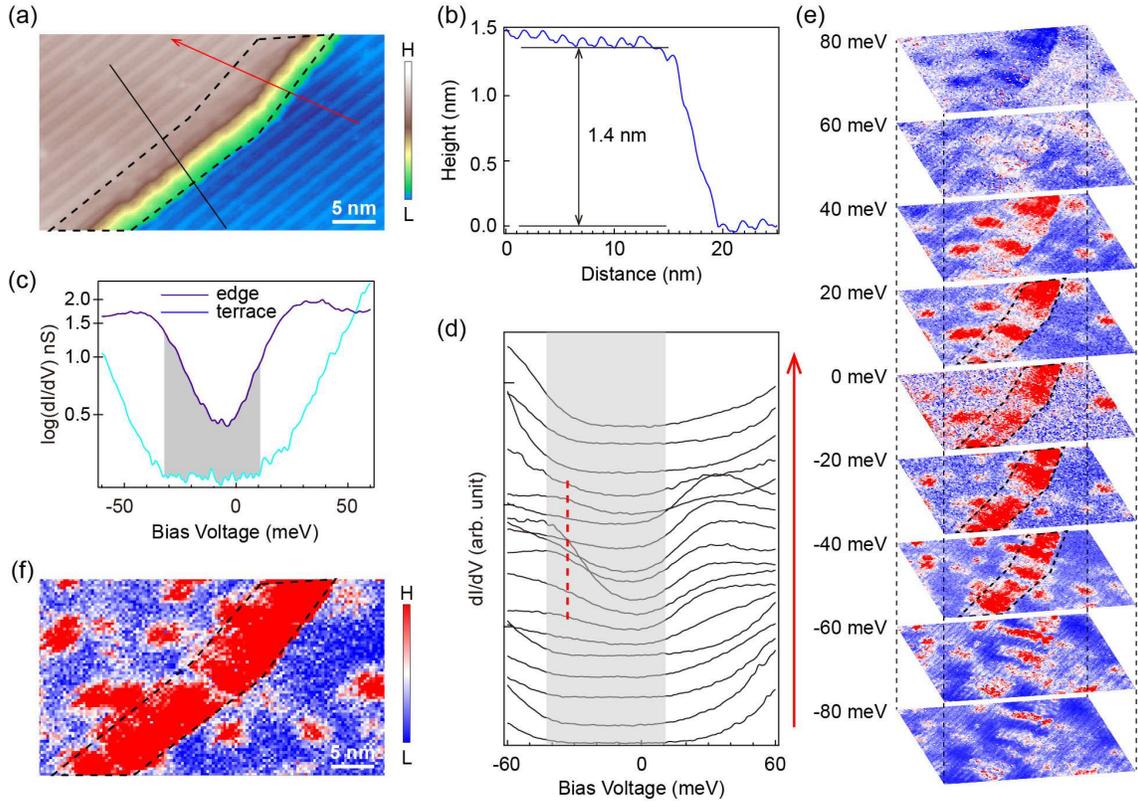}
\caption{{\bf STM characterization of the topological edge states in Ta$_2$Pd$_3$Te$_5$.} (a) STM topographic image containing a step edge. Dashed black lines indicate the position of the step edge. (b) Line profile along the solid black line in (a). (c) dI/dV curves taken on the flat terrace (blue) and near the step edge (black). (d) Series of dI/dV spectra taken along the red arrow in (a). The red dashed line indicates the emergence of in-gap states near the step edge. (e) dI/dV maps taken in the same area as (a). The bias voltage is indicated on the left side of each map. Pronounced edge states appear when the bias voltage is in the range of -40 to 20 meV, as indicated by the black dashed lines. (f) Averaged dI/dV map at four different bias voltages: -40, -20, 0, and 20 mV.}
\end{figure*}

Now that we have shown the existence of a topological band structure and a band gap in Ta$_2$Pd$_3$Te$_5$, we proceed to studying the topological edge states, which are a key signature of the QSH state in monolayer Ta$_2$Pd$_3$Te$_5$. Because of the weak interlayer coupling, topological edge states are expected to exist at the periphery of the topmost layers. An ideal technique to study the edge states is STS because the tunneling conductance is proportional to the local density of states (LDOS). Figure 3(a) shows an STM image that contains a step edge. From the line profile in Fig. 3(b), the step height is 1.4 nm, which corresponds to the lattice constant along the $a$ direction. Figure 3(c) shows the dI/dV curves taken on the flat terrace and near the step edge. On the flat terrace, we observe a band gap at the Fermi level, in agreement with the semiconductor behavior of Ta$_2$Pd$_3$Te$_5$. The estimated gap size is approximately 43 meV, with the valence band top and conduction band bottom at -33 meV and 10 meV, respectively. The band gap shows negligible variation across the flat terrace, despite the height variation of the Te chains, as shown in Supplementary Fig. S1. This indicates the high spatial homogeneity of the surface, which provides further evidence for the global nature of the band gap. Notably, the gap estimated from our STS data is larger than that fitted based on the transport measurements. This may originate from the surface sensitivity of the STS technique, which indicates a larger band gap in monolayer Ta$_2$Pd$_3$Te$_5$ than in the bulk material. Near the step edge, however, the LDOS is dramatically enhanced, featuring a V-shape in the energy range between -40 and 40 meV. This indicates the existence of edge states inside the band gap. To better visualize the evolution of the edge states, we present a series of dI/dV curves across the step edge in Fig. 3(d). When the tip approaches the step edge, the tunneling conductance inside the gap gradually increases, as indicated by the red dashed line in Fig. 3(d).

To confirm the spatial distribution of the edge states, we performed real space dI/dV mapping, as shown in Fig. 3(e). When the bias voltage is set to be within the band gap (e.g., 0 and -20 mV), the tunneling conductance is dramatically enhanced near the step edge. The enhancement of the tunneling conductance vanishes at bias voltages outside the band gap, resulting in a uniform LDOS over the entire surface (e.g., at $\pm$80 mV). Figure 3(f) shows an averaged dI/dV map at several bias voltages that span the band gap, where the enhancement of the LDOS near the step edge can be clearly seen. The fact that these edge states are located inside the gap agrees well with their topological nature. Notably, the step edge is not straight and contains several different terminations. Nevertheless, edge states always exist, despite the slight variation in the details of the STS spectra. This provides strong evidence for the robustness of the edge states, which is also consistent with their topological nature \cite{NieS2020}. Similar topological edge states have been reported in several other topological materials, such as ZrTe$_5$ \cite{WuR2016} and TaIrTe$_4$ \cite{DongX2019}.

In summary, our results support the existence of significant inverted gap and topological edge states in the van der Waals material Ta$_2$Pd$_3$Te$_5$, thus providing strong evidence for QSH states in Ta$_2$Pd$_3$Te$_5$. In stark contrast to WTe$_2$, multilayer and even bulk Ta$_2$Pd$_3$Te$_5$ hosts a similar inverted gap at the Fermi level. This is beneficial for device applications because multilayer samples are typically more inert and have higher tunability. Therefore, we expect that Ta$_2$Pd$_3$Te$_5$ will become a promising material for fabricating QSH devices.

\section*{Acknowledgments}
This work was supported by the Ministry of Science and Technology of China (Grants No. 2018YFE0202700, No. 2017YFA0302901, No. 2016YFA0300604, and No. 2016YFA0300904), the National Natural Science Foundation of China (Grants No. 11974391, No. 11974395, No. U2032204, No. 11774399, No. 11825405, and No. 1192780039), the Beijing Natural Science Foundation (Grants No. Z180007 and No. Z180008), the Strategic Priority Research Program of the Chinese Academy of Sciences (Grants No. XDB33030100 and No. XDB30000000), and the K. C. Wong Education Foundation (GJTD-2018-01). ARPES measurements were performed under the Proposal No. 19BG007. We thank the N-BARD, Hiroshima University for supplying liquid He.

\end{document}